# Electrically Tunable Polarizer Based on Graphene-loaded Plasmonic Cross Antenna

**Yuwei Qin [1,2], Xiaoyan Y.Z. Xiong [1], Wei E.I. Sha [1,3], and Li Jun Jiang [1]**

1. Department of Electrical and Electronic Engineering, the University of Hong Kong, Hong Kong.
2. Department of Electrical and Computer Engineering, Carnegie Mellon University, Pittsburgh, PA, 15213, USA.
3. Key Laboratory of Micro-nano Electronic Devices and Smart Systems of Zhejiang Province, College of Information Science & Electronic Engineering, Zhejiang University, Hangzhou 310027, China.

E-mail: yuweiq@andrew.cmu.edu, xyxiong@eee.hku.hk, weisha@zju.edu.cn and jianglj@hku.hk

**Abstract.** The unique gate-voltage dependent optical properties of graphene make it a promising electrically-tunable plasmonic material. In this work, we proposed in-situ control of the polarization of nanoantennas by combining plasmonic structures with an electrostatically tunable graphene monolayer. The tunable polarizer is designed based on an asymmetric cross nanoantenna comprising two orthogonal metallic dipoles sharing the same feed gap. Graphene monolayer is deposited on a Si/SiO$_2$ substrate, and inserted beneath the nanoantenna. Our modelling demonstrates that as the chemical potential is incremented up to 1 eV by electrostatic doping, resonant wavelength for the longer graphene-loaded dipole is blue shifted for 500 nm (~ 10% of the resonance) in the mid-infrared range, whereas the shorter dipole experiences much smaller influences due to the unique wavelength-dependent optical properties of graphene. In this way, the relative field amplitude and phase between the two dipole nanoantennas are electrically adjusted, and the polarization state of the reflected wave can be electrically tuned from the circular into near-linear states with the axial ratio changing over 8 dB. Our study thus confirms the strong light-graphene interaction with metallic nanostructures, and illuminates promises for high-speed electrically controllable optoelectronic devices.

## 1. Introduction

Dynamic manipulation of light at nanoscale is highly important in modern technology, and the demands promote the emergence of plasmonic nanoantennas, which convert light from free space into subwavelength volumes. Surface plasmons (SPs) due to the strong electron gas oscillation within the metallic nanoantennas facilitate flexible operations of light waves in many degrees of freedom including resonant frequency, bandwidth, radiation patterns and density of states [1]. Among all degrees of freedom, the polarization has drawn keen interests in recent years for its illuminating potentials in field enhanced microscopy and spectroscopy [2, 3], optical data storage [4], biochemical sensing [5], optical tweezers [6, 7] and nanophotonic integrated circuits [8]. Recently, designs featuring symmetric and asymmetric cross nanoantennas have been proposed as polarizers to manipulate the polarization of reflected beams [9, 10]. Both designs comprise two perpendicularly configured dipole nanoantennas sharing a common feed gap, and the desired polarization state of reflected beam can be tailored by engineering the geometries of the dipole arms. While the abovementioned designs are remarkable and novel, they are not convenient to tune the polarization state of the reflected beam once the device has been fabricated, and therefore dynamic tuning becomes particularly appealing. Many tuning mechanisms have been proposed to address the issue, including thermal [11], mechanical [12] and circuitry [13] attempts. Among all, electrical tuning, which is based on incorporation of phase-change media [14-16], is most prevailing thanks to the on-chip integration of subwavelength photonic components with electronics.

Of all material candidates, graphene stands out due to its superior electrical and thermal conductivities, widely tunable electro-optical properties, material abundance, and good chemical resistance. Consisting of a two-dimensional (2D) monolayer of carbon atoms arranged in honeycomb lattice, graphene has a zero energy band gap [17, 18]. The conductivity of graphene is highly dependent on the chemical potential, and the zero-gap nature allows it to achieve high electron or hole concentration via the electrostatic doping [19, 20]. Moreover, due to the forbidden interband transitions by Pauli blocking, the doped graphene has been identified as a nanoplatform to support SPs at terahertz or mid-infrared regime (MIR) [21-23]. As a result, the optical properties of graphene, such as intraband dispersion and interband loss via electron-hole pair excitation, can be dynamically controlled by tuning the carrier concentration (or chemical potential) through the applied gate voltage. Therefore, graphene is often integrated into other plasmonic devices and served as an electrically tunable load [24, 25].

In this work, by combining plasmonic nanoantennas with the electrostatically tunable graphene monolayer, we propose an in-situ control of the polarization of the reflected beam from the structure. The tunable polarizer is designed based on an asymmetric cross nanoantenna comprising two orthogonal metallic dipoles sharing the same feed gap. Graphene is deposited in between dipole antennas and Si/SiO$_2$ substrate. The strong light-graphene interaction, together with coupling with metallic structures, enables a broad tuning of the antenna resonances, which leads to the phase and polarization modifications for the reflected beam. Our simulation results demonstrate a broad tuning range over 500 nm for the graphene-loaded dipole in the mid-infrared regime by electrically doping the graphene chemical potential from 0 up to 1 eV. The tuning not only redistributes the field amplitudes along the antenna axes, but also adjusts the relative phase difference between the two perpendicular dipoles. Therefore, by measuring the axial ratio and Stokes parameters, we confirm that the reflected beam can be dynamically tuned from the circular to near-linear polarizations via the electrostatic doping of the graphene.

The rest of this paper is organized as follows. In section 2, we briefly review the relation between the dipole resonance and phase response. This relation serves as the route for most polarizer designs featuring cross plasmonic antennas. In section 3, we will discuss the optical properties of graphene, and explore physical understanding for the tuning mechanism when the chemical potential of graphene is tuned by the applied gate voltage. Section 4 demonstrates the tuning behaviours when the graphene sheet is integrated with plasmonic dipoles. Finally, section 5 presents our design and examines the electrically tunable polarizer with the graphene-loaded cross nanoantennas.

## 2. Phase response of plasmonic nanoantennas

We first consider a linear dipole nanoantenna made of gold placed on a SiO$_2$ substrate as shown in **Figure 1(a)**. Resembling a harmonic oscillator, it is expected that the phase shift between the driving field and the plasmonic field of the nanoantenna reaches -90˚ at resonance. A quantitative assessment of this behaviour is obtained with a first set of simulations, where the excitation light beam is incident towards the dipole plane with the electric field linearly polarized along the antenna axis. The wavelength range of interests lies in MIR, within which the electrical permittivity of SiO$_2$ and gold are, respectively, ε = 2.25 and described by Brendel-Bormann model [26]. The geometry is theoretically studied by rigorously solving Maxwell's equations with the finite-difference time-domain (FDTD) method [27].

To observe the resonance behaviour and phase response, the normalized extinction cross section (ECS) [28] and phase shift with respect to the driving field are calculated and plotted in **Figure 1(b)**. The single peak of the normalized ECS reveals a resonance at the wavelength of 5.7 μm. The phase shift at the resonance is -90° as expected, and it approaches 0° (-180°) as the incident wavelength ($\lambda_{in}$) is far larger (smaller) than the resonant wavelength. Also, as the length of the dipole arm increases, the resonance energy of the antenna is red shifted. Therefore, the resonance behaviour and phase response of the plasmonic dipole antenna can be modified by a proper choice of the antenna geometry.

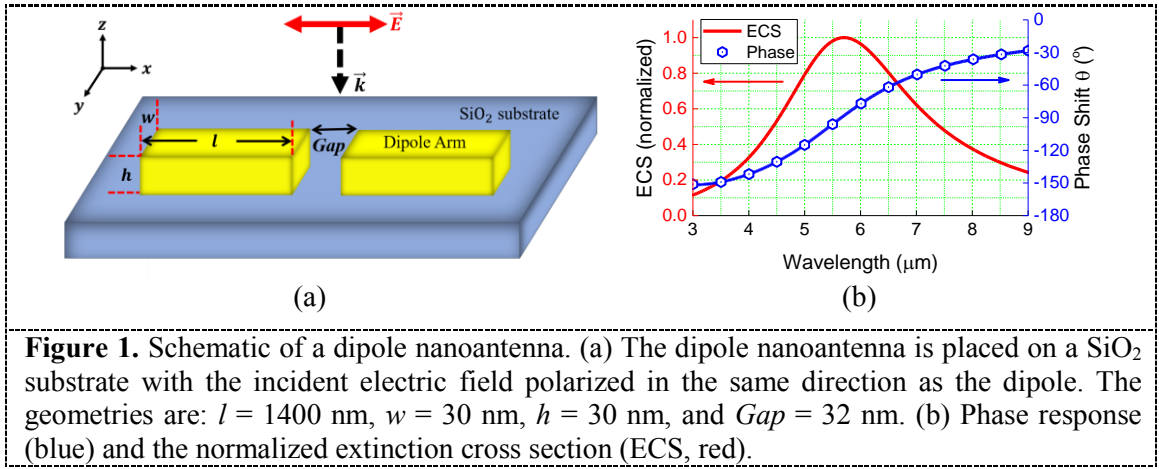

(a)          (b)

**Figure 1.** Schematic of a dipole nanoantenna. (a) The dipole nanoantenna is placed on a $SiO_2$ substrate with the incident electric field polarized in the same direction as the dipole. The geometries are: $l$ = 1400 nm, $w$ = 30 nm, $h$ = 30 nm, and *Gap* = 32 nm. (b) Phase response (blue) and the normalized extinction cross section (ECS, red).

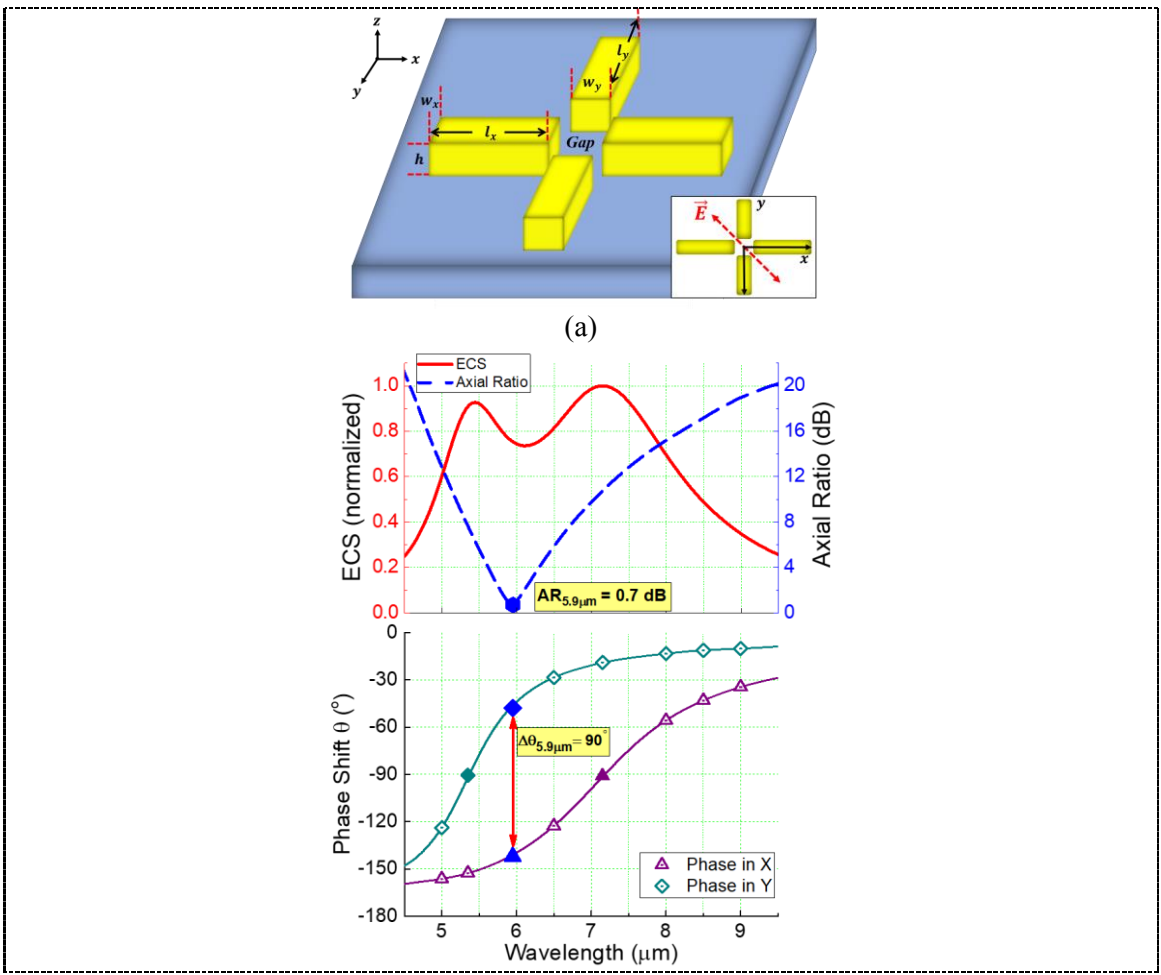

(a)

(b)

**Figure 2.** (a) Schematic of a static polarizer by a cross plasmonic antenna. The geometries are: [$l_x$, $l_y$] = [1400, 1100] nm, $w_x = w_y$ = 30 nm, $h$ = 30 nm, and $Gap$ = 32×32 nm$^2$. The inset shows the electric field of the incident beam. (b) Polarizer performance. Upper panel: normalized ECS (red) and axial ratio at far field (dashed blue). Lower panel: phase responses along both antenna axes.

Based on the resonance characteristics of the dipole nanoantenna, we now consider a static polarizer in **Figure 2(a).** This structure consists two plasmonic dipoles arranged perpendicular to each other with a mutually shared feed gap [9, 29]. Incident light beam is illuminated towards the dipole plane with the electric field linearly polarized at 45° with respect to the antenna axes shown as the inset. Upon illumination, both nanoantennas will enhance the field component parallel to its own axis. These two fields will then interact coherently in the feed gap with relative field amplitude and phase uniquely determined by the lengths of both dipoles. For example, a circularly polarized reflected beam at the operation wavelength of 5.9 μm is achievable if the antenna arm lengths are chosen at 1100 and 1400 nm, which are inferred from the single dipole phase-resonance relation. Under this configuration, both antennas yield roughly the same field enhancement as indicated from normalized ECS in the upper panel of **Figure 2(b)**. Furthermore, their respective phase shifts are -49° and -139° as described in the lower panel, thus creating a 90° phase difference overall. To characterize the polarization state of reflected beam, the axial ratio (AR) at far field, defined in equation (1), is taken as the figure of merit,

$$AR = 20 \log \left( \frac{E_{max}}{E_{min}} \right) \quad (1)$$

where $E_{max}$ and $E_{min}$ refer to the electric field amplitudes respectively at the major and minor axes of the polarization ellipse [30]. The result is plotted as the blue curve in the upper panel of **Figure 2(b)**, where AR indeed reaches the minimal value of 0.7 dB at the wavelength of 5.9 μm, confirming that the reflected beam is highly circularly polarized. On the other hand, if $\lambda_{in}$ is chosen in a way such that either the field amplitudes along both antenna axes become very unbalanced, or the net phase difference is far from 90°, the resulting polarization state will be much more elliptical. The key parameters in the working principle of this type of static polarizer are constituent nanoantenna geometries, which become completely inflexible once the structure is fabricated. To enable dynamic control, graphene is introduced as a tunable load.

### 3. Graphene optical properties and tuning mechanism

One of the main reasons why graphene has attracted considerable attention in plasmonic applications is that its optical properties are tightly connected with the chemical potential, which can be tuned by an external gate voltage. Surface conductivity of graphene monolayer is calculated with the Kubo formula [31, 32] in equation (2) as a function of temperature ($T$) of 300 K, frequency ($\omega$), carrier scattering rate ($\Gamma$) and chemical potential ($\mu_c$). In the calculation, the carrier scattering rate equals 11 meV/$\hbar$ [32, 33].

$$\sigma(\omega, \mu_c, \Gamma, T) = \frac{je^2}{\pi \hbar^2 (\omega - 2j\Gamma)} \left[ \mu_c + 2k_B T \log \left( -\frac{\mu_c}{k_B T} + 1 \right) \right] - \frac{je^2}{\pi \hbar} \log \left( \frac{2|\mu_c| - (\omega - 2j\Gamma)\hbar}{2|\mu_c| + (\omega - 2j\Gamma)\hbar} \right) \quad (2)$$

When the photon energy is larger than 2$\mu_c$, optical conductivity is dominated by the interband transition. As carrier concentration is gradually incremented due to the electrostatic doping such that the photon energy is less than 2μ$_c$, intraband transitions become more prominent while interband transitions are suppressed due to the Pauli blocking [21, 25, 34].

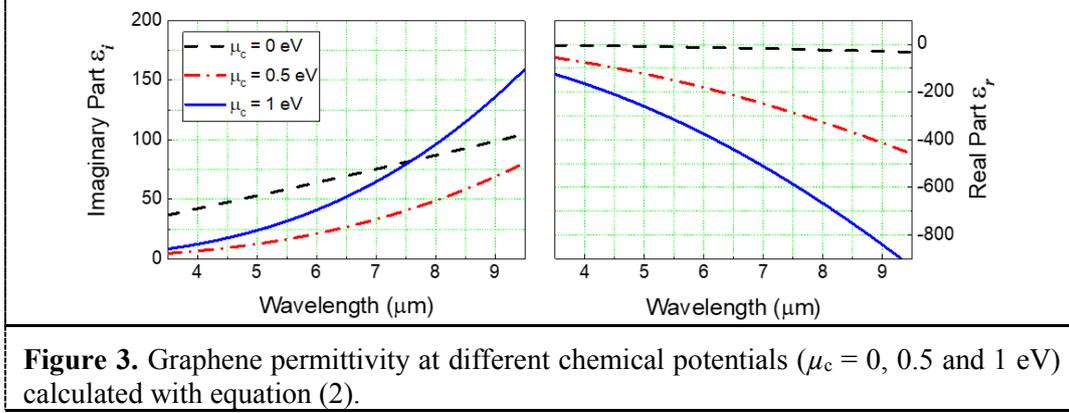

**Figure 3.** Graphene permittivity at different chemical potentials ($\mu_c$ = 0, 0.5 and 1 eV) calculated with equation (2).

The in-plane permittivity of graphene is defined as $\varepsilon_P @ \varepsilon_r - j\varepsilon_i = 1 - j\sigma/\omega\varepsilon_0 d_0$, where $d_0$ refers to the effective thickness of graphene sheet [22, 23], and σ is calculated by equation (2). Both real ($\varepsilon_r$) and imaginary ($\varepsilon_i$) components of the graphene in-plane permittivity are plotted in **Figure 3** for three different μ$_c$ values, while the out-of-plane permittivity is chosen to be the dielectric permittivity ($\varepsilon_\perp$ = 2.5) [22]. The real part of the in-plane permittivity, corresponding to the imaginary part of the graphene conductivity in equation (2), is reduced to negative values in MIR especially under high doping concentrations. Similar to noble metals, this phenomenon indicates that the highly doped graphene is capable of supporting transverse magnetic (TM) surface plasmon polaritons (SPPs), thus shining light upon strong light-matter interactions in graphene loaded nanostructures.

When graphene is introduced in the vicinity of plasmonic nanoantennas, especially in the feed gap, the resonance behaviour is dependent on the graphene doping level. Given the fact that graphene monolayer is extremely thin, perturbation approach [22, 35] is taken here to gain insight on the tuning mechanism. For simplicity, consider Maxwell's equations in the eigenmodes where electric field is expressed in Dirac notation, $|E\rangle$, with a time convention of e$^{-i\omega t}$. Given a source-free linear system, Maxwell's equation reads

$$\nabla \times \nabla \times |E\rangle = \left(\frac{\omega}{c}\right)^2 \varepsilon |E\rangle \quad (3)$$

The perturbation from graphene monolayer is accounted for as the order parameter Δε, representing the change of permittivity. In this way, the first-order correction of resonant frequency ω$^{(1)}$ is determined by expanding the eigenmode and matching the order parameter Δε as:

$$\omega^{(1)} = -\frac{\omega^{(0)}}{2} \frac{\langle E^{(0)} | \Delta\varepsilon | E^{(0)} \rangle}{\langle E^{(0)} | \varepsilon | E^{(0)} \rangle} \quad (4)$$

where ω$^{(0)}$, ε and $|E^{(0)}\rangle$ represent the resonance, permittivity and eigenmode without the perturbation. Qualitatively, equation (4) shows that as Δε grows more negative due to a higher doping concentration of graphene monolayer, the resonance of the graphene-loaded plasmonic antenna will be blue shifted.

## 4. Graphene-loaded plasmonic dipole antenna
To quantitatively demonstrate the tuning capacity of graphene when combined with metallic nanostructures, two graphene-loaded dipole structures are considered in **Figure 4(a)**. In type I structure, the graphene nanoribbon (GNR) is inserted beneath the gold dipole arms, while in type II, the GNR only covers the gap area and does not spatially overlap the antenna arms. The thickness of GNR is assumed 0.35 nm in the modelling with 7 layers of meshing grids along the *z*-coordinate, and the chemical potential is incremented from 0 up to 1 eV in response to the external gate voltage change.

To illustrate resonance tuning, the normalized ECS's for type I structure at different chemical potentials are presented in **Figure 4(b)**. As the chemical potential increments, the in-plane permittivity of graphene drops drastically (**Figure 3**), representing a decreased perturbation $\Delta\varepsilon$ in equation (4). Hence, tuning the chemical potential of graphene to a higher level leads to a growing first order correction to the resonant frequency ($\omega^{(1)}$), and thus significantly blue shifts the resonance of the graphene-loaded dipole. Furthermore, it is worth mentioning that the introduction of GNR, even at the zero doping level ($\mu_c = 0$), leads to a blue shift of 80 nm in comparison to the unloaded dipole. This is a result of the peculiar linear dispersion relation of graphene at $\mu_c = 0$ [21, 33, 36], which gives rise to the negative permittivity ($\varepsilon_r$) even at the zero doping.

Because of highly enhanced light intensity in the feed gap [9, 22, 37, 38], the light-graphene interaction is most concentrated within the nano-gap. As a proof, **Figure 4(c)** compares the tuning ranges at each chemical potential for both type I and II configurations. As expected, as long as the graphene covers the entire gap area, there is no preferred direction in which the strong graphene-light interaction is taken place, and the tuning capacity for both structures are essentially the same at all tested doping levels.

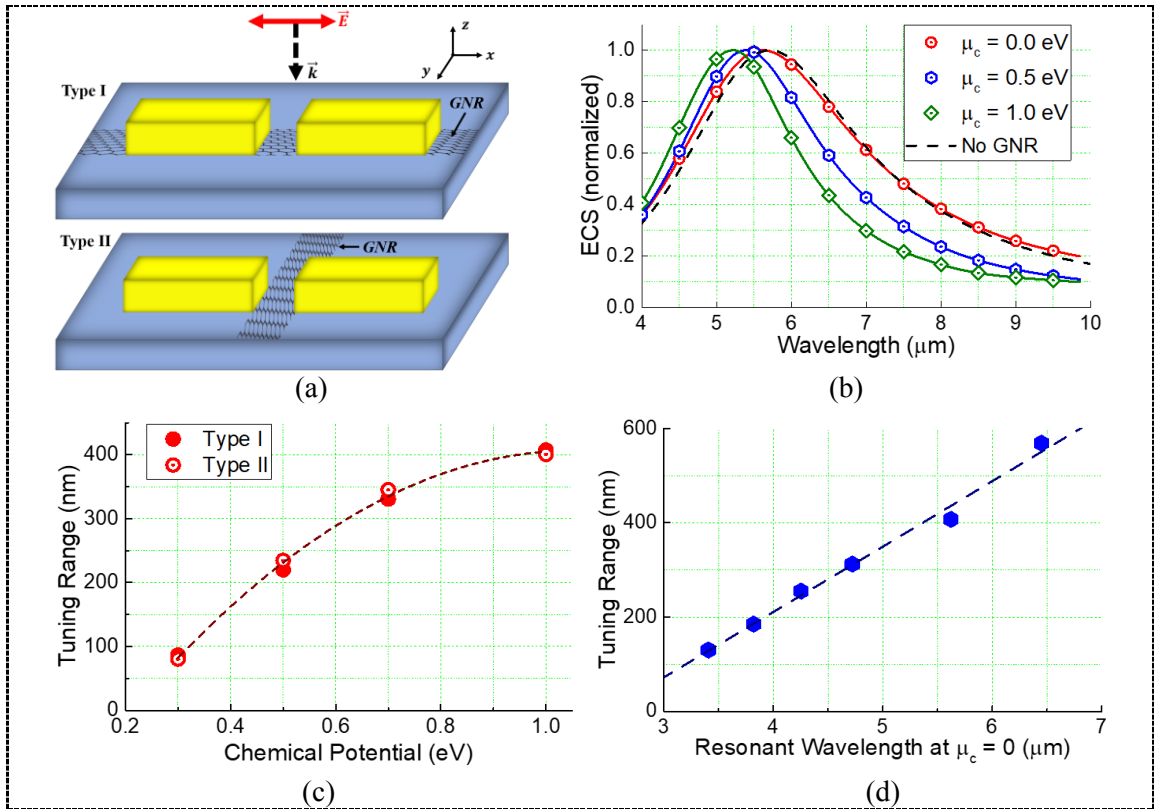

**Figure 4.** Graphene-loaded dipole structures. (a) Schematics of type I and II graphene-loaded dipoles where GNR is deposited beneath (I) and perpendicular to (II) the dipole arms. Antenna geometries are identical to those in **Figure 1(a)**. (b) Resonance tuning for type I dipole at different chemical potential levels. The resonance wavelengths are: [$\lambda_{\text{no GNR}}$, $\lambda_{\text{0eV}}$, $\lambda_{\text{0.5eV}}$, $\lambda_{\text{1eV}}$] = [5710, 5630, 5400, 5210] nm. (c) Tuning range evaluated as a function of chemical potentials for type I (solid) and II (dotted) structures. (d) Tuning range ($|\lambda_{\text{1eV}} - \lambda_{\text{0eV}}|$) plotted as a function of the undoped graphene-antenna resonant wavelength ($\lambda_{\text{0eV}}$). The dashed lines in (c) and (d) are the fitting curves.

**Figure 4(d)** presents the tuning range for the type I structure as a function of the undoped graphene-antenna resonant wavelengths, which is modified by adjusting the dipole arm lengths. This relation is crucial because the resonance tuning due to geometric difference is at the heart of the plasmonic polarizer design as discussed in section 2. It is clear from the figure that as the resonant wavelength increases, tuning range grows linearly as a natural consequence of the enlarged in-plane permittivity amplitude ($|\varepsilon_r|$). If the operation wavelength is chosen at a fairly small value, 3.8 μm for example, the tuning range is only 180 nm, whereas at a longer operation wavelength (6.4 μm), the tuning range could reach 580 nm, approximately 9% shift of the resonance.

## 5. Electrically tunable polarizer design

Based on the phase-resonance relation and the graphene-loaded dipole tuning principle, we design an electrically tunable polarizer as shown in **Figure 5**, which sketches both the whole structure (left) and a single unit cell (right). The proposed design modifies the traditional static polarizer by introducing a graphene monolayer in between the gold cross nanoantenna and Si/SiO$_2$ substrate. It is important to maintain the feed gap small in order to enhance the field intensity within the gap and thus promoting the field-graphene overlap [22]. Three electrodes are introduced, namely source, drain and gate, in order to enable externally electrostatic doping of graphene monolayer. The first two electrodes have direct contacts with the graphene monolayer, and the gate is laid at the bottom beneath substrate. During operation, the drain and source are maintained at the same voltage while a variable bias is applied to the gate. In this way, the graphene-substrate-gate structure demonstrates similar characteristics as a capacitor, and enables to electrostatically dope the graphene monolayer.

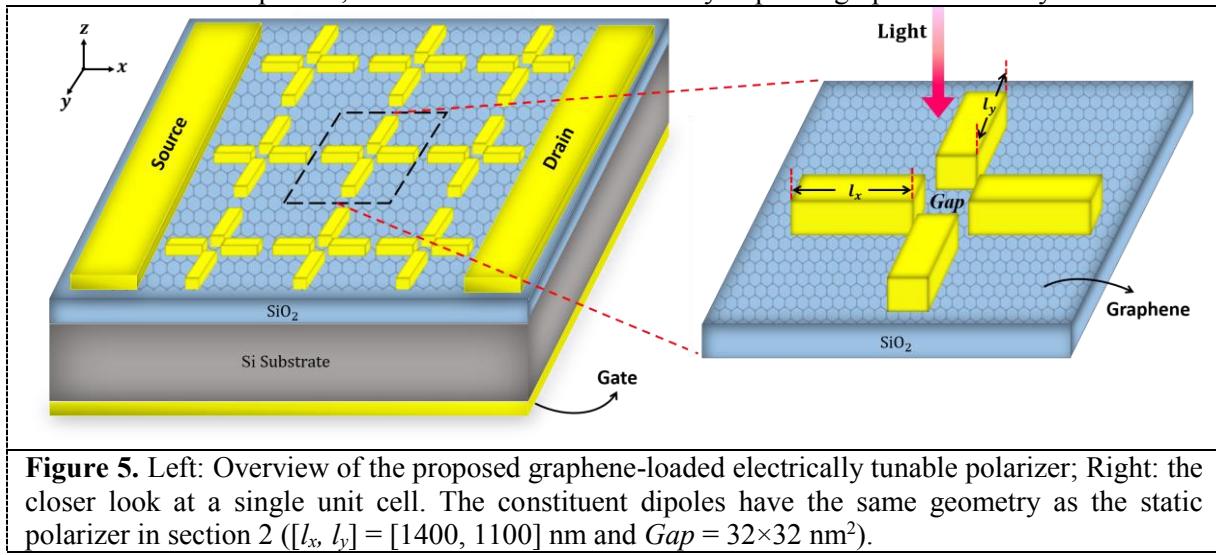

**Figure 5.** Left: Overview of the proposed graphene-loaded electrically tunable polarizer; Right: the closer look at a single unit cell. The constituent dipoles have the same geometry as the static polarizer in section 2 ($[l_x, l_y]$ = [1400, 1100] nm and $Gap$ = 32×32 nm$^2$).

The principle of the tunable polarizer is described as follows. At zero doping level, the constituent dipole geometries are arranged such that at the operation wavelength $\lambda_0$, the reflected beam is circularly polarized with balanced electric field components along both antenna axes and a 90° phase difference. As the gate voltage is adjusted to increase the chemical potential of graphene monolayer, the resonance responses of both dipoles are blue shifted. However, since the longer dipole with a longer resonant wavelength exhibits more significant or sensitive response to the chemical potential tuning as explained in the section 4, the reflected beam thus results in unbalanced electric field intensities along both the antenna axes. Furthermore, the relative phase difference between the reflected electric field from both dipoles will no longer maintain at 90°. Altogether, the unbalanced electric field intensities along the antenna axes and the adjusted relative phase difference tune the polarization of the reflected beam off the circular state.

The polarizer performance for a unit cell is demonstrated in **Figure 6** where the graphene chemical potential is increased from 0 to 1 eV. Since graphene is covering the entire feed gap, resonances for both the dipoles are blue shifted in **Figure 6(a)**, and the shifted wavelengths are determined by the antenna geometries. In our case, the dipole along the *x*-coordinate is longer than its counterpart, therefore the resonance shift of the second peak is more significant than that of the first one. Axial ratios at far field for both the chemical potentials are plotted in **Figure 6(b)**. Within wavelengths of interests, a minimal AR is achieved for the graphene-loaded polarizer with and without doping at 6 and 5.5 µm. They mark the upper and lower bounds of the incident light wavelengths within which a highly circularly polarized reflected beam can always be reached by adjusting the chemical potential of graphene monolayer with the gate voltage. On the other hand, if the wavelength of the incident light beam is predetermined and fixed, tuning the graphene doping level naturally enables a dynamic control of the reflected beam polarization. For example, at 6 µm operation wavelength, the undoped polarizer generates almost a perfect circular state of the reflected beam with 0.3 dB of AR. If the doping level is then increased to 1 eV, AR is enlarged to almost 9 dB, indicating highly elliptical or linearly polarized state depending on the required threshold.

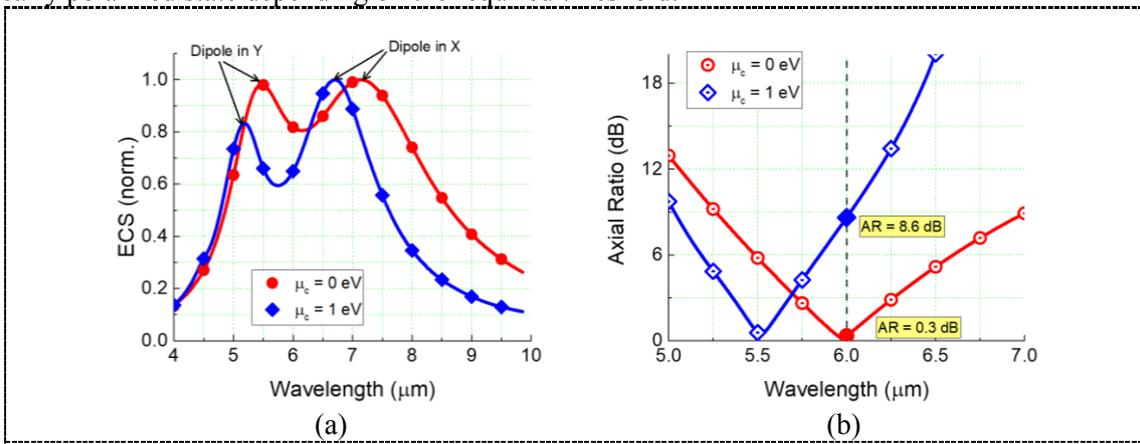

(a)                                    (b)

**Figure 6.** (a) Normalized ECS for graphene-loaded polarizer with a zero doping (red) and chemical potential of 1 eV (blue). The two peaks located at the wavelength of 5.3 and 5.5 µm correspond to the resonances of the dipole in the *y*-coordinate ($l_y$ = 1100 nm), and those located at 6.7 and 7.2 µm are for the dipole in the *x*-coordinate ($l_x$ = 1400 nm). (b) Axial ratio (AR) of the reflected beam for both chemical potentials.

Another support to this polarization change is by evaluating Stokes parameters at near field [39]. Stokes parameters are 4×1 vectors where the first entry (S0) describes the total intensity of the light beam, the second and third (S1, S2) differentiate linear polarization, and the last (S3) characterizes the level of circularly polarized state. **Figure 7** plots the normalized S1 and S3 at the operation wavelength of 6 µm concentrating the feed gap region. At the zero doping, the Stokes parameters evaluated within the feed gap are dominated by the value of S3 (**Figure 7(b)**), confirming the reflected beam is highly circular. If the chemical potential is tuned to a higher value, the blue-shifted resonance of the longer dipole (*x*-coordinate) becomes more adjacent to the operation wavelength. Consequently, the reflected beam is more linearly polarized along the *x*-direction as confirmed in (**c, d**).

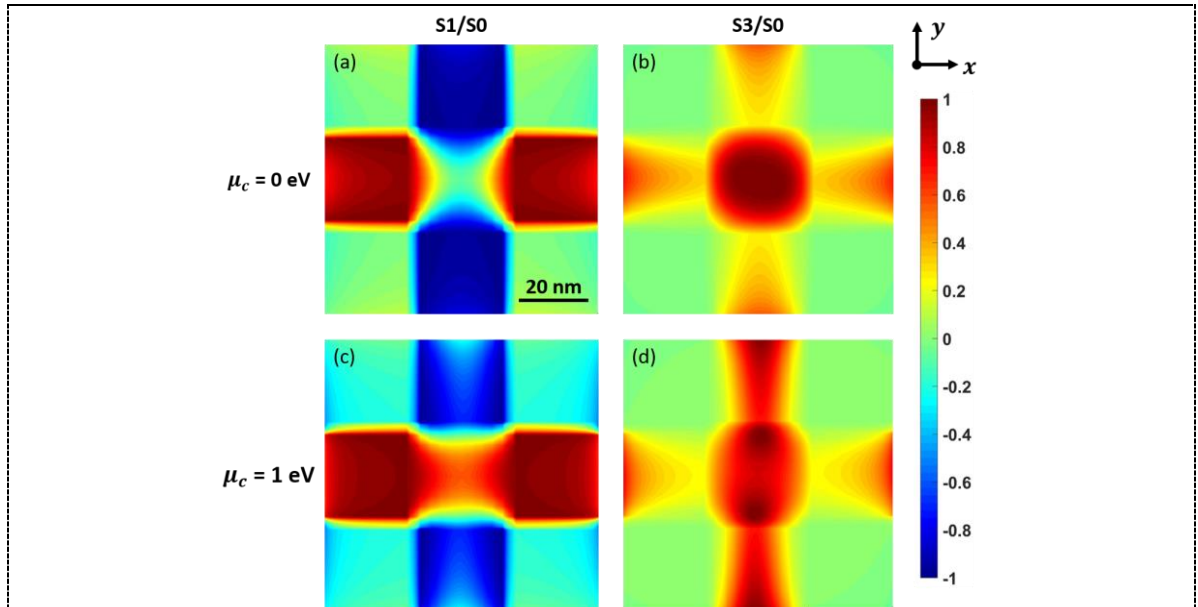

**Figure 7.** Stokes parameters calculated at 20 nm above the polarizer top surface focusing on the polarizer feed gap. The operation wavelength is 6 μm with the graphene chemical potential varied between 0 (first row) and 1 eV (second row). The first column plots S1/S0, and second column S3/S0. Stokes vectors measured at the center of the feed gap are $\vec{S}_{0eV}$ = [1, 0.0, 0.1, 1.0]$^T$ and $\vec{S}_{1eV}$ = [1, 0.6, 0.1, 0.8]$^T$, where $^T$ denotes transpose.

## Conclusion

To conclude, in this work we have analysed the tuning behaviour of graphene at the MIR when integrated with other plasmonic nanostructures. Despite atomic thickness, graphene monolayer has demonstrated significant tuning capacity especially when the interacting field is highly localized and enhanced. Combing a graphene monolayer with a cross nanoantenna, we design an electrically tunable polarizer which enables in-situ control over the polarization of the reflected beam. Our simulations have demonstrated over 500 nm tuning range for a highly circularly polarized reflected beam, and more than 8 dB axial ratio change at far field. The results illuminate the capability of electrically manipulating light polarization at nanoscale with graphene loads, and shine light upon applications featuring polarization control.

## Acknowledgments

This work was supported in part by the Research Grants Council of Hong Kong (GRF 17207114 and GRF 17210815), NSFC 61271158, Hong Kong UGC AoE/P–04/08, and and Hundred Talents Program of Zhejiang University (No. 296 188020*194231701/208). The author would like to thank Dr. Xingang Ren and Miss Menglin Chen for their valuable inputs in modelling nanostructures.